\documentstyle[12pt]{article}
\input{epsfig.sty}
\newcommand{\beq}{\begin{eqnarray}}
\newcommand{\eeq}{\end{eqnarray}}
\begin{document}
\title{Dark Mass Creation During EWPT Via Dark Energy Interaction}
\author{Leonard S. Kisslinger and Steven Casper\\
Department of Physics, Carnegie Mellon University, Pittsburgh, PA 15213}
\date{}
\maketitle

\begin{abstract}
  We add Dark Matter Dark Energy terms with a quintessence field interacting
with a Dark Matter field to a MSSM EW Lagrangian previously
used to calculate the magnetic field created during the EWPT.
From the expectation value of the quintessence field we estimate the Dark
Matter mass for parameters used in previous work on Dark Matter-Dark Energy
interactions.

\end{abstract}
\maketitle
\noindent
PACS Indices:12.15Ji,12.60.Cn,98.80.Cq,95.35.+x

\noindent
Keywords: Cosmology; Electroweak Phase Transition; Dark Matter; Dark Energy

\section{Introduction}

  Our present work is based on a Minimal Supersymmetry Model (MSSM) of
the Electroweak (EW) Lagrangian which was used to calculate electromagnetic
field creation during nucleation\cite{hjk10} and magnetic field creation
during bubble collisions\cite{sjkhhb08} during the Electroweak Phase 
Transition (EWPT) that occurred at a time $t=10^{-11}$ seconds, when the 
critical temperature was $T_c$ =125 GeV. In the present work we add terms to 
the Lagrangian for the Dark Energy quintessence field and the interaction of a 
Dark Matter field with the quintessence field, based on models introduced in 
Refs\cite{pr88,fp04}.

  In the MSSM EW theory the EWPT is first order, so there is critical 
temperature and bubbles of the new universe form within the old universe. The 
latent heat for the EWPT is the value of the Higgs field, $\Phi$, which goes 
from $<\Phi>=0$ to $<\Phi>=v \simeq$ 125 GeV when $T=T_c$. At this time the 
Higgs gets a mass $M_H$, as do all particles in the standard model except the 
photon:
\beq
\label{massesduringewpt}
               M_H &=& v  \nonumber \\
               M_W &=& g v/\sqrt{2} {\rm\;\;g\;=\;strong\;coupling\;constant} \\
               M_Z &=& M_W/cos(\theta_W) {\rm\;\;\theta_W\;=\;Weinberg\;angle}
\nonumber \\
         m_e&\propto&m_u \propto m_d \propto v \nonumber \; .
\eeq

  In the Standard Model
\beq
\label{weakms}
           M_W&=& 37/sin(\theta_W) \simeq 80 {\rm \;GeV} \nonumber \\
           M_Z &\simeq& 90 {\rm \;GeV} \; .
\eeq

  There have been a number of studies of the origin of Dark Matter mass.
If the EWPT is first order, which it is in our MSSM theory, baryogenesis
occurs, with more particles than antiparticles. One model of Dark Matter 
mass generation unifies Dark Matter and baryogenesis\cite{kmss09}. See
this reference for references to earlier related publications. More recently
a study using Ref\cite{pr88} for the quintessence field derived Dark Matter
mass in terms of mass varying neutrinos\cite{cank11}.
  
  Since all standard model particles got their mass during the EWPT, our
present work is based on the hypothesis that Dark Matter also got its
mass during the EWPT via interaction with the quintessence field; and we use 
the techniques developed in Refs\cite{hjk10,sjkhhb08,pr88,fp04} to carry out 
the calculation of Dark Matter mass.
\section{MSSM EW equations of motion with quintessence field}

  We add to the MSSM Lagrangian used earlier to study electromagnetic
field creation\cite{hjk10} and magnetic field creation \cite{sjkhhb08}  
additional terms for the Dark Energy quintessence field and with the 
interaction of the quintessence field with the Dark Matter Fermion field,
from which we calculate the Dark Matter mass.
\beq
\label{L}
  {\cal L}^{MSSM} & = & {\cal L}^{1} + {\cal L}^{2}  + {\cal L}^{3} 
+ {\cal L}^{fermion} +{\cal L}^{DM-DE} \\
     {\cal L}^{1} & = & -\frac{1}{4}W^i_{\mu\nu}W^{i\mu\nu}
  -\frac{1}{4} B_{\mu\nu}B^{\mu\nu}  \nonumber \\
 {\cal L}^{2} & = & |(i\partial_{\mu} -\frac{g}{2} \tau \cdot W_\mu
 - \frac{g'}{2}B_\mu)\Phi|^2  -V(\Phi) \nonumber \\
 {\cal L}^{3} &=& |(i\partial_{\mu} -\frac{g_s}{2} \lambda^a C^a_\mu)\Phi_s|^2
    -V_{hs}(\Phi_s,\Phi) \nonumber \\
    {\cal L}^{fermion}& = & {\rm \;standard \;Lagrangian \;for \;fermions}
\nonumber \\
    {\cal L}^{DE}&=& \frac{1}{2}\partial_\nu \Phi_q \partial^\nu \Phi_q
-V(\Phi_q) \nonumber \\
    {\cal L}^{DM-DE}&=& g_D \bar{\psi}^{DM} \Phi_q \psi^{DM} \nonumber \; .
 \nonumber \; .
\eeq

  A cosmological constant was introduced\cite{guth} to produce
inflation in the very early universe, which solved the problem of our
homogeneous universe as shown by the observation of temperature correlations
in the Cosmic Microwave Background Radiation. The quintessence field  $\Phi_q$
was used to model inflation, see Ref\cite{dv02}. In this model $\Phi_q$ vanishes
at a very early time, but recent studies of supernova velocities and galaxies
show that dark energy is now about 3/4 of the matter in the universe. We
use the model of Refs\cite{pr88,fp04} with the present dark energy being
created at the time of the EWPT.
\newpage

Following Refs\cite{pr88,fp04},
\beq
\label{Vphi}
      V(\Phi_q)&=& K \Phi_q^{-\alpha} \; ,
\eeq
where $K$ and $\alpha$ are parameters which must be chosen. 

This gives the
differential equation for $\Phi_q$, neglecting ${\cal L}^{DM-DE}$
\beq
\label{phieq}
             \partial^2 \Phi_q -K \alpha \Phi_q^{-(\alpha+1)}&=& 0 \; .
\eeq

  For $K$ we use the value in Ref\cite{pr88}, for $\alpha$ use the range
$2.0-6.0$ given in Refs\cite{pr88,fp04}.  Ferrar and Peebles\cite{fp04}
showed that the preferred range is $4.0\le \alpha \le 6.0$, so we give
our results for this range of $\alpha$ in seperate figures.

The scale of the universe, $a(t)$, is defined as $a(t)=R(t)/R_o$, where
$R(t)$ is the radius of the universe at time $t$ and $R_o$ is the radius
at the present time. The solution to Eq(\ref{phieq}) for $\Phi_q(t)$ is
\beq
\label{phisoln}
    \Phi_{EWPT}&\simeq&[2 \alpha(\alpha+2)]^{1/2} (\frac{a(t_{EWPT})}
{a(t_1)})^{3/(\alpha+2)} \; ,
\eeq
with $\Phi_q(t_{EWPT})$ the quintessence field at the time of the EWPT
and $t_1>>t_{EWPT}$  is to be chosen.
 
Making use of the solutions of the General Theory of Relativity, the radius
of the universe has a time dependence in a radiation dominated universe
$R(T)\propto t^{1/2}$. Therefore with $t_{EWPT}=10^{-11}s$ and $t_1$ in seconds,
\beq
\label{ratio}
        \frac{a(t_{EWPT})}{a(t_1)}&=&\sqrt{\frac{10^{-11} s}{t_1}} \; .
\eeq

We use the model of Ref\cite{pr88}, with the dark matter mass, $M_{DM}$,
given in our theory with $t$ the time of the EWPT, and ${\cal L}^{DM-DE}$:
\beq
\label{DMeq}
        M_{DM}&=&g_D \frac{m_p}{32 \pi} \Phi(t_{EWPT}) \; .
\eeq

Since the Planck mass $m_p =1.22\times10^{19}$ GeV, from 
Eqs(\ref{DMeq},\ref{phisoln},\ref{ratio}), and using\cite{KT} $g_D=\pi \times
10^{-11}$
\beq
\label{massDM}
   M_{DM}&=& 3.82\times10^6 [2 \alpha(\alpha+2)]^{1/2}(\sqrt{\frac{10^{-11} s}
{t_1}})^{3/(\alpha+2)} \; .
\eeq 

For $t_1$ we use both $t_{eq}$=1,500 years, when the universe went from being 
radiation dominated to matter dominated, which is consistent with
the theory in Refs{\cite{pr88,fp04}, and $t_{now}$=13.7 billoion years, in
which scenerio the Dark Energy field evolved until the present time. 

  Using Eq(\ref{massDM}) we calculate the Dark Matter Mass for these two
final times. The results are shown in the figures.
\clearpage

The solutions for $M_{DM}$ for $t_1$=$t_{eq}$=1,500 years
with the values of $\alpha$ expected\cite{pr88,fp04}
are shown in Figure 1:
 
\vspace{8cm}

\begin{figure}[ht]
\begin{center}
\epsfig{file=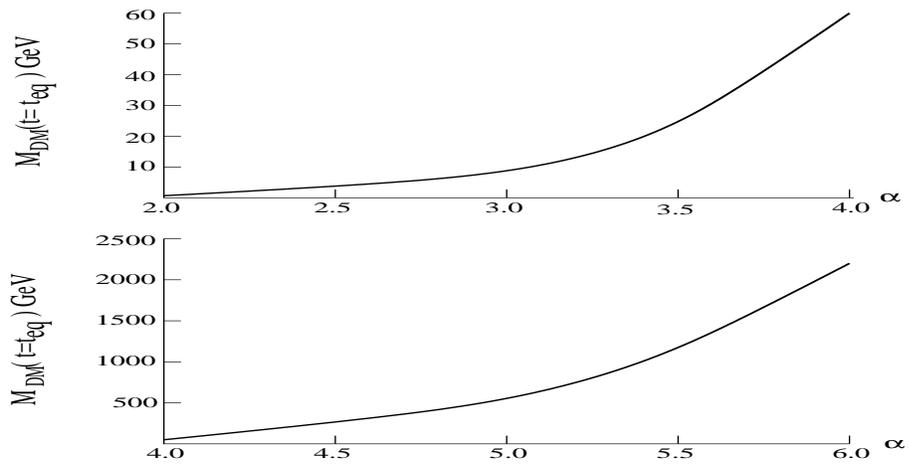,height=6cm,width=12cm}
\caption{$M_{DM}$ for $t_1=t_{eq}$ for $2.0\le \alpha \le 4.0$ and the
preferred values $4.0\le \alpha \le 6.0$ }
\label{Figure 1}
\end{center}
\end{figure}
\clearpage
The solutions for $M_{DM}$ for $t_1$=$t_{now}$=$13.7 \times 10^9$ years
with the values of $\alpha$ expected\cite{pr88,fp04}
are shown in Figure 2:
 
\vspace{8cm}

\begin{figure}[ht]
\begin{center}
\epsfig{file=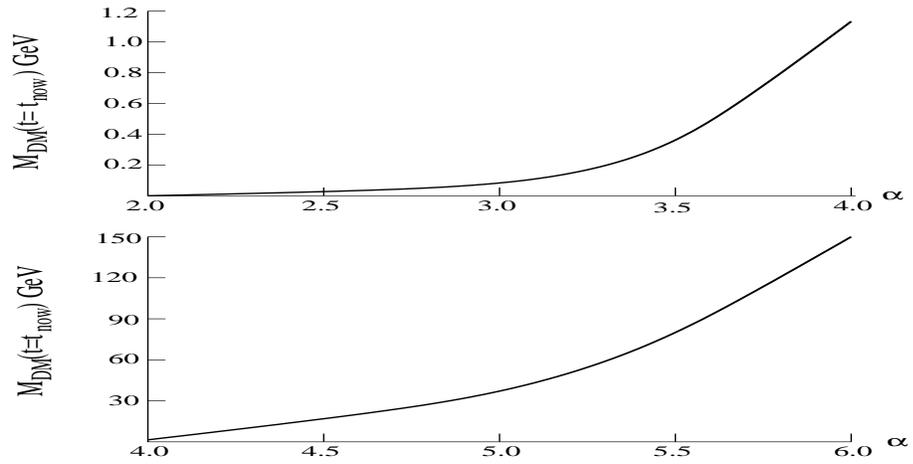,height=6cm,width=12cm}
\caption{$M_{DM}$ for $t_1=t_{now}$ for $2.0\le \alpha \le 4.0$ and the
preferred values $4.0\le \alpha \le 6.0$}
\label{Figure 2}
\end{center}
\end{figure}
\vspace{2cm}

\newpage
 
\section{Conclusion}
  We have derived the Dark Matter mass using the concepts that since all
standard particles got their masses during the EWPT from interaction with
the Higgs field, it would be consistent for Dark Matter, which has only a 
gravitational force, to get its mass starting from the time of the EWPT 
via interaction with the dark energy (quintessence) field.

  Using the solution for the quintessence field in Ref\cite{pr88} with the
MSSM EW Lagrangian from Ref\cite{hjk10}, the Dark Matter masses have been
derived for the  range of the parameter $\alpha$ from 2.0 to 6.0. For the 
most prefered values of $\alpha$\cite{pr88} from 4.0 to 6.0 for the 
final time $t_1=t_{eq}$, which is the most appropriate as the universe went 
from radiation to matter dominated,
the Dark Matter masses that have been found are from about 100 GeV to 2 TeV, 
which is consistent with values that have been predicted. If we use 
$t_1=t_{now}$, the predicted value for the Dark Matter masses for values of  
$\alpha$ from 4.0 to 6.0 go from a few GeV to 140 GeV, which are smaller than
expected.

 Therefore we conclude that Dark Matter might have obtained its mass via
interaction with the Dark Energy field during the EWPT, just as the standard
model particles got their masses via interaction with the Higgs field at
that time.

\end{document}